\documentclass[prb,amsmath,amssymb,twocolumn,superscriptaddress,showpacs]{revtex4-1}
\usepackage{bm}
\usepackage{amssymb}
\usepackage{colordvi}
\usepackage{graphicx}
\usepackage{color}
\usepackage{hyperref}
\usepackage{ulem}
\usepackage{color}

\newcommand{\beq}{\begin{equation}}
\newcommand{\eeq}{\end{equation}}
\newcommand{\bea}{\begin{eqnarray}}
\newcommand{\eea}{\end{eqnarray}}

\newcommand{\nn}{\nonumber}
\newcommand{\vep}{\varepsilon}

\def\Im {\mbox{Im}}

\def\nn{\nonumber}

\begin{document}

\title{Band topology in classical waves: Wilson-loop approach to topological numbers and fragile topology}

\affiliation{School of Physical Science and Technology, Guangxi Normal University, Guilin 541004, China}
\affiliation{College of Physics, Optoelectronics and Energy, \&
	Collaborative Innovation Center of Suzhou Nano Science and
	Technology, Soochow University, 1 Shizi Street, Suzhou 215006,
	China}
\affiliation{Department of Physics and Center for Theoretical Physics, National Taiwan University, Teipei 10617, Taiwan}
\affiliation{Physics Division, National Center for Theoretical Sciences, Hsinchu 30013, Taiwan}

\author{Hai-Xiao Wang}
\affiliation{School of Physical Science and Technology, Guangxi Normal University, Guilin 541004, China}
\affiliation{Department of Physics and Center for Theoretical Physics, National Taiwan University, Teipei 10617, Taiwan}
\affiliation{Physics Division, National Center for Theoretical Sciences, Hsinchu 30013, Taiwan}

\author{Guang-Yu Guo}
\affiliation{Department of Physics and Center for Theoretical Physics, National Taiwan University, Teipei 10617, Taiwan}
\affiliation{Physics Division, National Center for Theoretical Sciences, Hsinchu 30013, Taiwan}

\author{Jian-Hua Jiang}\email{jianhuajiang@suda.edu.cn}
\affiliation{College of Physics, Optoelectronics and Energy, \&
  Collaborative Innovation Center of Suzhou Nano Science and
  Technology, Soochow University, 1 Shizi Street, Suzhou 215006,
  China}


\begin{abstract}
The rapid development of topological photonics and acoustics calls for accurate understanding of band topology in classical waves, 
which is not yet achieved in many situations. Here, we present the Wilson-loop approach for exact numerical calculation of the 
topological invariants for several photonic/sonic crystals. We demonstrate that these topological photonic/sonic crystals are 
topological crystalline insulators with fragile topology, a feature which has been ignored in previous studies. We further 
discuss the bulk-edge correspondence in these systems with emphasis on symmetry broken on the edges.
\end{abstract}

\maketitle

\section{\label{sec:level1}Introduction}

The concept of band topology was introduced into physics since the seminal work by Thouless, Kohmoto, Nightingale, and den
Nijs~\cite{Thouless1982}. The integer quantum Hall effect is characterized by the total Chern number
${\cal C}$ of all bands below the Fermi level~\cite{Thouless1982,KOHMOTO1985343,Haldane1988}. The Hall conductance is
related to this topological number by the rigorous relation $\sigma_{xy}={\cal C} e^2/h$. The past decades has witnessed the 
rise of various topological phases of electrons in condensed matter systems, particularly, topological insulators, semimetals and
superconductors~\cite{Hasan2010,Qi2011,tsm1,kanemele,bhz,konig,fukane1,fukane2,fukane3,ckxu,jiang2,murakami,burkov2011weyl,jiang1,luling1}. 
In recent years, the study of band topology in classical waves, such as electromagnetic
waves~\cite{c1,c2,c3,c4,poo2011,hafezi2013imaging,Mittal2016,rechtsman2013photonic,Khanikaev2013,ma2015guiding,chen2014experimental,Wu2015,Xu2016,deformedPhC,Wang2016,c10,c11,c12,z2new,wang2017type,Lu2014,Lu2016,RMPtopophoton}, 
acoustic waves~\cite{Yang2015topoacousitc,Xiao2015,He2016,chen2014,wuying,va1,va2,liu3,zhu1,zhu2,Zhang2017,Jiang2018}, and mechanical waves~\cite{mechanical1,mechanical2,mechanical3,z2linenode}
with and without time-reversal symmetry, has attracted lots of attention, due to their fundamental distinction from electrons
as well as the emergent, unprecedented ways of molding the flow of the classical waves governed by topological mechanisms.
Differing from electrons, there is no Fermi level and band filling is not an issue for classical waves (i.e., classical waves can
be excited in a very broad band of frequencies). Topological numbers in classical waves
are associated with band gaps (or partial band gaps), which consist of contributions from all bands below the gap.

To date, several prototypes of topological band-gaps and partial band-gaps have been proposed or observed in a number of 
different systems including photonic and sonic crystals. However, in much of the existing studies, topological numbers are obtained
 from analog with electronic systems, rough arguments, or indirect evidences (such as band inversion and edge states).
In this work, we present systematic calculations of the Zak phase, Chern number,  {and Wannier bands} for various one-dimensional (1D) and two-dimensional (2D) topological photonic and sonic crystals using first-principle calculation. Through these calculations, the topological nature of the band-gaps in those photonic and acoustic systems are clarified unambiguously. Furthermore, we also reveal the physics of ``fragile topology" in classical wave systems that mimic topological
crystalline insulators protected by space groups. In particular, we use sonic crystals with $C_6$ symmetry as an example to 
show that the Wilson-loop approach is an ideal tool to study fragile topology in classical wave systems. The Wilson-loop
approach adopted in this work is reviewed in details for acoustic and photonic systems. We further discuss the bulk-edge
correspondence in those photonic and sonic crystals by considering realistic boundary conditions and possible symmetry 
broken mechanisms on the edges.

The remaining of the paper is organized as follows: in Sec.~\ref{sec:level2} we present the theory of Wilson-loop approach 
in photonic and acoustic systems. This theory is derived from the basic definition of Berry phases and topological numbers 
in photonic and acoustic systems. We also present numerical schemes for the implementation of the Wilson-loop calculation 
in these systems. In Sec.~\ref{sec:level3} we apply the method to several 1D and 2D photonic and acoustic systems, as 
concrete examples. We give discussions on the bulk-edge correspondence and its possible issues for these systems in 
Sec.~\ref{sec:level4} and then conclude in Sec.~\ref{sec:level5}.

\section{\label{sec:level2}Formalism}

\subsection{Berry connection and Berry phases for classical waves}
Topological numbers characterize the global topological properties of the Bloch wavefunctions of the energy bands. 
These energy bands are defined on the wavevector space: an ``infinite" noncompact space for continuous media or systems, 
while the finite compact Brillouin zone (BZ) for periodic systems. The study in this work is restricted to periodic systems such 
as photonic and sonic crystals. Periodic systems are usually much more complicated (but also more versatile) than continuous 
systems. In the latter case the topological number can often be calculated analytically. It is known
that the topological numbers of a band can be obtained by integration over various Berry phases in the BZ. Direct integration of
Berry phase encounters gauge ambiguity and is inconsistent with numerical schemes since continuous evolution of the Bloch 
wavefunctions in the BZ are demanded for such calculations. Another approach, the Wilson-loop approach, is gauge invariant 
and compatible with numerical implementations. We will derive the Wilson-loop approach for electromagnetic and 
acoustic waves in periodic systems. We then apply this approach for the study of topological numbers in photonic and sonic crystals.

Before we introduce the Wilson-loop approach in details, the first thing to notice is that the formulation of Berry phase and Berry
connection in classical waves are fundamentally different from the formulation of these quantities in electronic systems. 
Let us start with electromagnetic waves. A concrete derivation of the Berry phase for electromagnetic waves was established
in Ref.~\onlinecite{Onoda2006}. Without magneto-electric coupling, the Berry connection in electromagnetic wave is given by~\cite{Onoda2006}
\beq
\vec{\Lambda}_{n,{\vec k}}=\tfrac{1}{2} [\vec{\Lambda}_{n,{\vec k}}^{\it E} + \vec{\Lambda}_{n,{\vec k}}^{\it H}]
\eeq
where
\begin{subequations}
	\begin{align}
		{\vec \Lambda}_{n,{\vec k}}^{\it E}&=\frac{{\rm i}  \int d{\vec r}  {\vec E}_{n,{\vec k}}^{\dagger}({\vec r}) \cdot \hat{\vep}({\vec r})\cdot  \nabla_{{\vec k}}{\vec E}_{n,{\vec k}}({\vec r})}{\int d{\vec r}  {\vec E}_{n,{\vec k}}^{\dagger}({\vec r}) \cdot \hat{\vep}({\vec r})\cdot  {\vec E}_{n,{\vec k}}({\vec r}) },\\
		{\vec \Lambda}_{n,{\vec k}}^{\it H}&=\frac{ {\rm i}\int d{\vec r} {\vec H}_{n,{\vec k}}^{\dagger}({\vec r})\cdot \hat{\mu}({\vec r})\cdot \nabla_{{\vec k}}{\vec H}_{n,{\vec k}}}{ \int d{\vec r} {\vec H}_{n,{\vec k}}^{\dagger}({\vec r})\cdot \hat{\mu}({\vec r})\cdot {\vec H}_{n,{\vec k}}},
	\end{align}
\end{subequations}
are the contributions from the electric and magnetic fields, respectively. Here ${\vec E}_{n,{\vec k}}({\vec r})$ and 
${\vec H}_{n,{\vec k}}({\vec r})$ are the periodic part of the vectorial Bloch wavefunctions of the $n$-th photonic band with wavevector 
${\vec k}$ for the electric and magnetic fields, respectively. Here the symbol $\dagger$ denotes complex conjugation and transpose 
for vectors and matrices. $\hat{\vep}({\vec r})$ and $\hat{\mu}({\vec r})$ give the spatial dependences of the permittivity and permeability 
tensors, respectively. Throughout this paper, the integration over ${\vec r}$ is performed within a unit-cell. 

Since the electric and magnetic fields can be determined from one another, one of them is enough to determine the 
Berry connection. In other words, $\Lambda_{n,{\vec k}}^{\it E}=\Lambda_{n,{\vec k}}^{\it H}$. However, there may be situations with 
magneto-electric coupling, where we need a more general form for the Berry connection of electromagnetic Bloch wavefunctions. It is 
proven in Ref.~\onlinecite{z21} that the general form is given by
\beq
{\vec \Lambda}_{n,{\vec k}} = \frac{{\rm i} \int d{\vec r} {\vec U}_{n,{\vec k}}^{\dagger}({\vec r})\cdot \hat{M}({\vec r})\cdot \nabla_{{\vec k}}{\vec U}_{n,{\vec k}}({\vec r})}{\int d{\vec r} {\vec U}_{n,{\vec k}}^{\dagger}({\vec r})\cdot \hat{M}({\vec r})\cdot {\vec U}_{n,{\vec k}}({\vec r})} .
\eeq
Here ${\vec U}_n^T=({\vec E}_{n}^T, {\vec H}_{n}^T)$, and $\hat{M}({\vec r})$ is the position-dependent $6\times 6$
electromagnetic susceptibility tensor which is written as
\beq
\hat{M} = \left( \begin{array}{ccc}
	\hat{\vep} & \hat{\xi} \\
	-\hat{\xi} & \hat{\mu}
\end{array} \right) . \label{MO}
\eeq
where the $3\times 3$ tensor $\hat{\xi}$ describes the magneto-electric coupling. For simplicity, we have ignored the 
frequency-dependence of the tensor $\hat{M}$. A more careful treatment can be directly established from the above formulation 
using the expression for the Berry connection of dispersive materials presented in Ref.~\onlinecite{z21}. Physical constraints on 
the susceptibility tensor and their consequences on the Berry phase is also analyzed in Ref.~\onlinecite{z21}.

We now define the ``inner product" of the Bloch wavefunctions as follows,
\beq
\langle  \psi_{n,{\vec k}}({\vec r})|\psi_{n,{\vec k}^\prime} ({\vec r})\rangle \equiv ({\cal N}_{n,{\vec k}}{\cal N}_{n,{\vec k}^\prime})^{-\frac{1}{2}}\int d{\vec r} \psi_{n,{\vec k}}^{\dagger}({\vec r})\cdot \hat{{\cal L}}({\vec r})\cdot \psi_{n,{\vec k}^\prime} ({\vec r})
\eeq
where $\psi_{n,{\vec k}}({\vec r})$ is the ${\vec k}$-dependent ``wavefunction" of the classical wave, $\hat{{\cal L}}({\vec r})$ is 
energy density operator which makes the integration on the right hand side be proportional to the energy of the classical wave, when 
${\vec k}={\vec k}^\prime$. The pre-factors on the right hand side is defined as,
\beq
{\cal N}_{n,{\vec k}} = \int d{\vec r} \psi_{n,{\vec k}}^{\dagger}({\vec r})\cdot \hat{{\cal L}}({\vec r})\cdot \psi_{n,{\vec k}}({\vec r}) .
\eeq
In this way, the inner product is normalized. The Berry connection for the $n$-th band is then defined as
\beq
{\vec \Lambda}_{n,{\vec k}} = {\rm i} \langle  \psi_{n,{\vec k}}({\vec r})|\nabla_{{\vec k}}\psi_{n,{\vec k}}({\vec r})\rangle . \label{bn}
\eeq
The above form applies to both electromagnetic and acoustic waves. For electromagnetic waves, the position dependent 
energy density operator is $\hat{{\cal L}}({\vec r})=\hat{M}({\vec r})$, as given in Eq.~(\ref{MO}). For acoustic waves, 
the energy density operator is $\hat{{\cal L}}({\vec r})=\frac{1}{2\rho({\vec r}) v({\vec r})^2}$, where $\rho$ is the mass density, 
$v$ is the speed of sound (both of them depends on the medium at the position ${\vec r}$). Therefore, the normalization factor 
for an acoustic Bloch wavefunction (rigorously speaking, the periodic part of the Bloch function), $p_{n,{\vec k}}({\vec r})$, which 
gives the acoustic pressure distribution for the $n$-th acoustic band with wavevector ${\vec k}$, is given by
\beq
{\cal N}_{n,{\vec k}} = \int d{\vec r} p_{n,{\vec k}}^{\ast}({\vec r})  \frac{1}{2\rho({\vec r}) v({\vec r})^2} p_{n,{\vec k}}({\vec r})  .
\eeq

With the above formalism for the Berry connection (\ref{bn}), the Berry phases can then be calculated whenever the system undergoes 
an adiabatic evolution along a closed loop. For instance, the Berry phase for adiabatic evolution of Bloch wavefunction along a 
closed loop ${\cal O}$ in the BZ is defined as
\beq
\theta_{n} = \oint_{{\cal O}} d{\vec k} \cdot {\vec \Lambda}_{n,{\vec k}} .
\eeq

\subsection{Wilson-loop approach}
We shall start with the calculation of topological number in 1D classical waves. In this situation, the topological number is denoted as
the Zak phase~\cite{Zak1989}, which is defined as
\beq
\theta=\int_{-\pi}^{\pi} dk \Lambda_{n, k}
\eeq
where we have scaled the length by the lattice constant, i.e., we set the lattice constant to be unity. In systems with inversion 
symmetry, Zak phase can only be 0 (trivial) or $\pi$ (nontrivial)~\cite{Zak1989}. The integral of the Berry connection over
the BZ $-\pi \le k < \pi$ can be obtained by dividing the BZ into many small segments and approximating the integral as the
summation of the contributions from each small segment. For instance, if the BZ is divided into $N$ segments, in a
small segment from ${k_i}$ to $k_{i+1}$ ($i$ range from 1 to $N$, and $k_{N+1}=k_1$), the Berry phase $\theta_i$ is given by
\begin{align}
	e^{-{\rm i}\theta_i}&= \langle \psi_{n,k_i}| \psi_{n,k_{i+1}} \rangle .
\end{align}
In the above we have used the fact that for a sufficiently small segment
\begin{align}
	e^{-{\rm i}\theta_i}& \approx 1 - {\rm i} \theta_i=1- {\rm i}\Lambda_{n, k}\delta
	k ,
\end{align}
and replaced the differentiation of Bloch wavefunctions over $k$ in $\Lambda_{n}$ by finite differences.

The total Berry phase is the summation of the Berry phase in each small segment,
\beq
\theta=\sum_{i=1}^N \theta_i .
\eeq
This direct numerical scheme, however, needs a continuous phase-fixing scheme for the Bloch wavefunctions at different 
$k$'s to ensure that the evolution of the Bloch wavefunctions is continuous in the BZ. As envisioned by Wilson, the total 
gauge phase can be calculated by accumulation of the Berry phase from each small segment in the following way
\beq
e^{-{\rm i}\theta}=\prod_{i=1}^{N} e^{-{\rm i}\theta_i} = \prod_{i=1}^{N}\langle \psi_{n,k_i}| \psi_{n,k_{i+1}}\rangle.
\eeq
Remarkably, the above formalism is gauge-invariant, since each Bloch wavefunction appears twice in the above product: one with
complex conjugation, the other with complex conjugation. Arbitrary determination of the phases of the Bloch wavefunctions, which 
often happens in numerical calculations, does not affect the calculated Berry phase $\theta$. This gauge-invariant
formalism is the basic starting point of the Wilson-loop approach.

The situation becomes more complicated when there are multiple energy bands. For such situations, in the Wilson-loop approach, 
the Berry phase for each segment is fully represented by the matrix $\hat{M}^{k_i,k_{i+1}}$ of which the matrix elements are given by
\begin{align}
	& M_{nn^\prime}^{k_i,k_{i+1}} = \langle U_{n, k_i}|
	U_{n^\prime,k_{i+1}}\rangle, \quad n, n^\prime \in 1...{\cal N}. \label{MultiBC}
\end{align}
where the band indices $n$ and $n^\prime$ go over all bands below the concerned band gap. If there are ${\cal N}$ bands 
below the concerned band gap, then the matrix is a ${\cal N}\times {\cal N}$ matrix. This matrix contains all
information on the Berry phase for the bands below the gap (or more precisely, this is the Berry connection matrix 
for the small segment $[k_i, k_{i+1}]$). From this matrix one can obtain all topological numbers of the bands below the gap,
e.g., the Chern number for quantum anomalous Hall effects. 

The Berry phase (in a matrix form) for a loop in the BZ can be obtained by the matrix product of the ${\cal N}\times {\cal N}$ Berry 
connection matrices for the small segments in the loop through the following form,
\beq
\hat{W} = \prod_{i=1}^N \hat{M}^{k_i,k_{i+1}} , \label{WN1}
\eeq
To evaluate the topological numbers, we need the eigenvalues of the above Berry phase matrix. The Berry phases associated with
the eigenvalues are given by
\beq
\theta_n \equiv {\rm i} \log(w_n) , \quad n=1,..., {\cal N},
\eeq
where $w_n$ is the $n$-th eigenvalue of the matrix $\hat{W}$. Since
$\theta_n$ must be a real number, the above can also be written as
\beq
\theta_n \equiv - \Im \log(w_n) , \quad n=1,..., {\cal N}. \label{WN2}
\eeq

The Chern number is related to the Berry phase through the following relations,~\cite{hmwengreview}
\begin{align}
	2\pi {\cal C}_n  & =-\int_{-\pi}^{\pi}\int_{-\pi}^{\pi}dk_x dk_y(\partial_{k_x}\Lambda^{(y)}_{n,{\vec k}}-\partial_{k_y}\Lambda^{(x)}_{n,{\vec k}})\nn\\
	& =\int_{-\pi}^{\pi} dk_y \partial_{k_y} [\int_{-\pi}^{\pi} dk_x \Lambda^{(x)}_{n,{\vec k}} ]\nn\\
	& \equiv \int_{-\pi}^{\pi} d\theta_{n,k_y},  
\end{align}
where
\begin{align} \label{WC}
	\theta_{n,k_y}  & \equiv \int_{-\pi}^{\pi} dk_x \Lambda^{(x)}_{n,{\vec k}} 
\end{align}
is the Berry phase for the $n$-th band along the loop $k_x\in [-\pi, \pi]$ for a fixed $k_y$ which is obtained by the integration 
over the Berry connection $\Lambda^{(x)}_{n,{\vec k}}$ (i.e., the $x$ component of ${\vec \Lambda}_{n,{\vec k}}$) through 
Eqs.~(\ref{WN1}) and (\ref{WN2}) for the Wilson-loop along $k_x$. 
In numerical calculations, the Chern number ${\cal C}_n$ is obtained by counting the winding phase of $\theta_{n,k_y}$ when 
$k_y$ goes from $-\pi$ to $\pi$. We remark that although the Berry phase $\theta_{n,k_y}$ depends on the choice of the origin of the 
spatial coordinates, the winding phase of $\theta_{n,k_y}$, however, does not depend on such choices. Thus, the Wilson-loop 
approach provides an effective and gauge-invariant method to evaluate the topological Chern number numerically. If there are 
multiple bands below the band gap, the total Chern number is given by the summation of the contributions from all bands 
below the concerned band gap, i.e., $2\pi {\cal C} = \sum_n \int_{-\pi}^{\pi} d\theta_{n,k_y}$.

 {Before proceeding, it is important to point out that the Berry phase in Eq.~(\ref{WC}) is related to
the Wannier center of the 2D crystal~\cite{dipolemoments}. For the case where the Berry phase is calculated at fixed $k_y$, the
Berry phase and Wannier center are related by $\nu_x = \theta/2\pi$ (Here, we set the lattice constant as unity; We may mix the
words ``Berry phase" and ``Wannier bands" in the remaining part of the paper). The evolution of 
the Wannier center with respect to $k_y$ is denoted as the ``Wannier bands". As has been shown in Ref.~{YuRui}, 
Wannier bands can be used to characterize
both the time-reversal broken Chern insulator and the time-reversal invariant quantum spin Hall insulator. The former
is illustrated above. The latter is characterized by gapless Wannier bands in half of the Brillouin zone~\cite{YuRui}, 
i.e., $k_y\in [0, \pi]$. Here, we shall also use the gapless Wannier bands as the signature of quantum spin Hall
effect for photons~\cite{Khanikaev2013,ma2015guiding,chen2014experimental,Wu2015,Xu2016,Wang2016,wang2017type,lu2016symmetry} 
and other classical waves~\cite{He2016,chen2014}.}

\section{\label{sec:level3}Numerical Results}

In this section, we calculate the topological numbers for several photonic/sonic crystals in 1D or 2D systems numerically by 
using the Wilson-loop approach elaborated above.

\subsection{\label{Zak calc}Zak phase in 1D photonic crystals}

\begin{figure}[b]
	\includegraphics[width=3.2in]{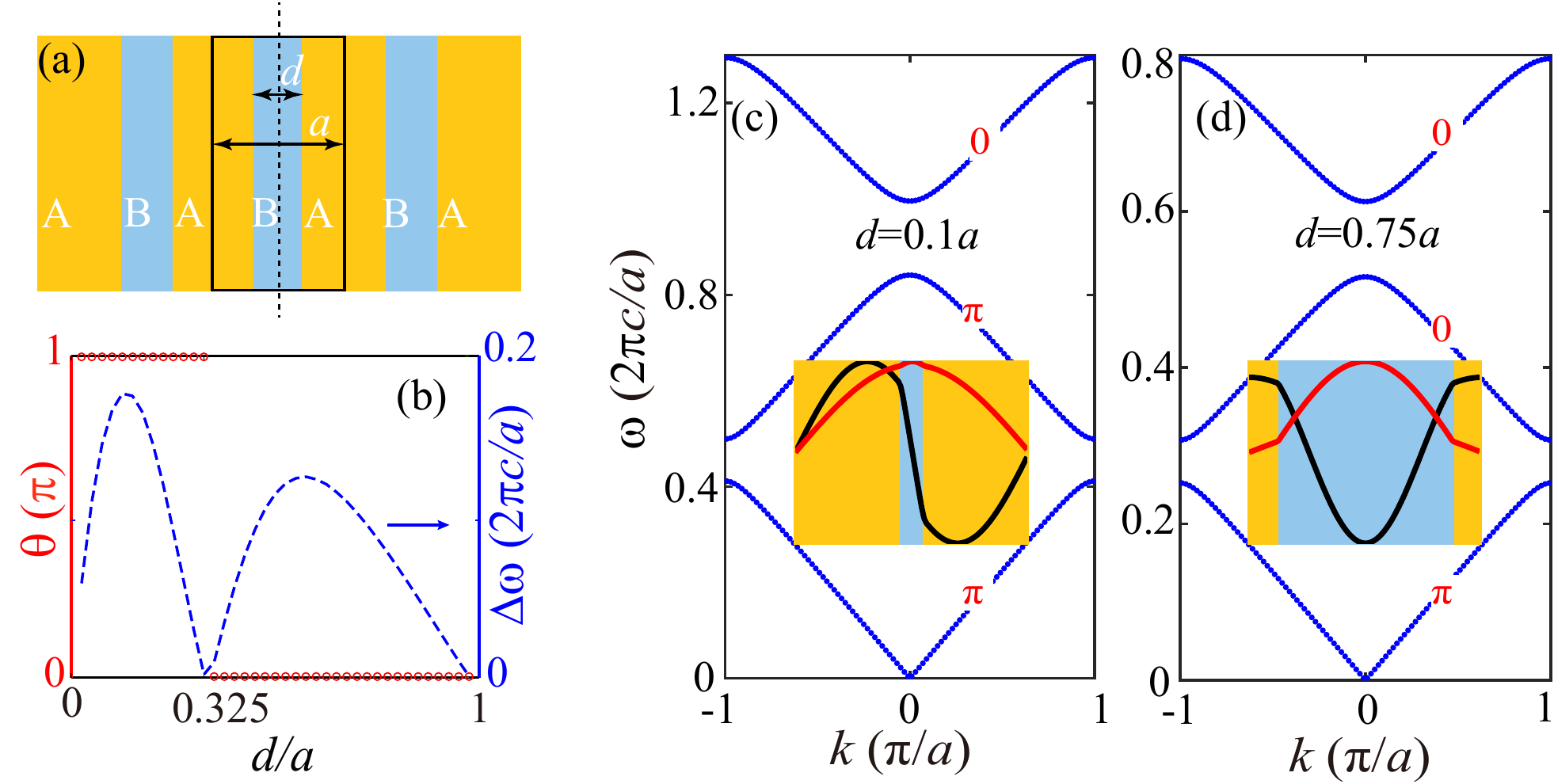}
	\caption{Numerical calculation of Zak phase in 1D photonic crystal. (a) Schematic of 1D photonic crystal consists of dielectric (orange area, $\epsilon_A$=4) and air (light blue area, $\epsilon_B$=1). The dotted line indicates the center of unit cell. All lengths are scaled by lattice constant $a$. (b) The Zak phase and the second band gap $\Delta\omega$ versus the thickness of dielectric. Red circles (dash line) refer to the Zak phase of the second band (band gap size). (c,d) The band structure of 1D photonic crystal with (c) $d=0.1a$ and (d) $d=0.75a$, respectively. The Zak phase of each individual band is labeled in red. Inset: the distribution of electric field with $k=0$ (black line) and $k=\tfrac{\pi}{a}$ (red line).}
\end{figure}

\begin{figure}[b]
	\includegraphics[width=3.4in]{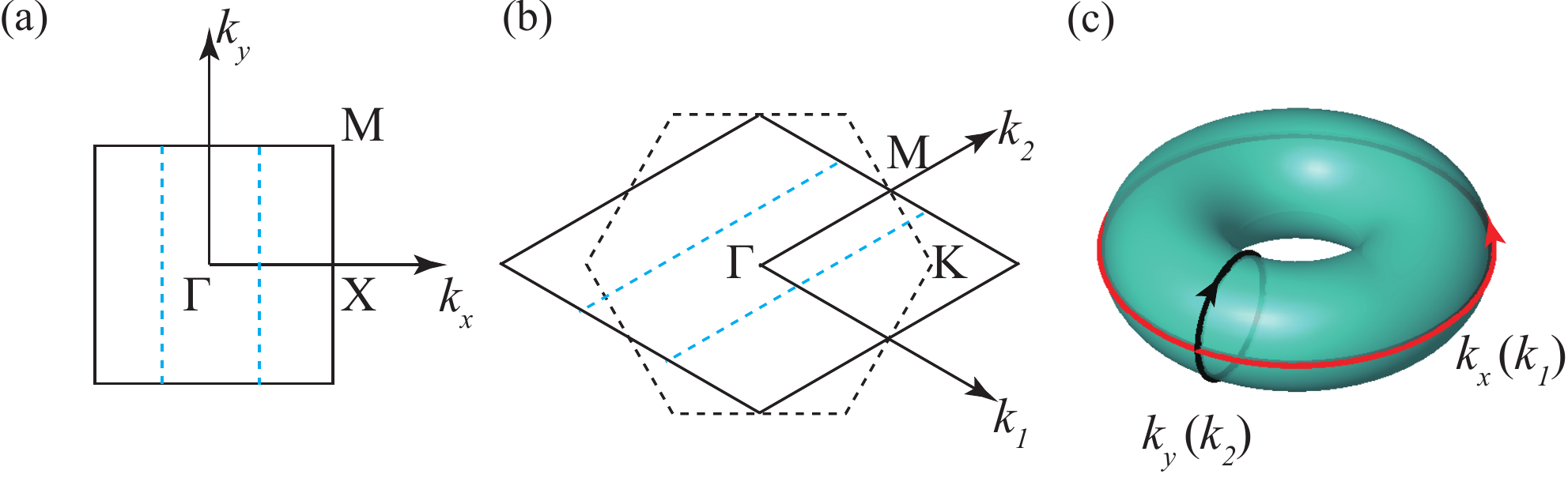}
	\caption{The equivalence of 2D BZ and torus. The BZ of (a) square lattice and (b) triangular lattice. (c) The 2D BZ is equivalent to a torus under the periodic boundary condition for the Bloch states.}
\end{figure}

We start with a 1D photonic crystal shown in Fig.~1(a) for transverse-electric(TE) modes. The material and geometric parameters are 
specified in Figure caption. The photonic band structures and the electromagnetic Bloch wavefunctions are calculated
numerically via the open software MIT PHOTONIC BANDS. The Zak phase of the second band is calculated using the 
Wilson-loop approach for various $d$ ($d$ is the thickness of the $B$ layer). The numerical Zak phase is indeed quantized 
to 0 or $\pi$, because the system has inversion symmetry~\cite{Zak1989}. The topological phase transition takes place at 
$d=0.325a$ where the Zak phase abruptly changes from $\pi$ to 0 [see Fig.~1(b)]. The photonic band gap closes at the 
transition point, where the second and third bands touch each other linearly, forming a 1D Dirac point.
To gain further insights, we plot the band structure and electric field profiles at the $k=0$ and $k=\pi$ points for two cases with 
$d=0.1a$ (topological) and $d=0.75a$ (trivial), respectively. For $d=0.1a$ [see Fig.~1(c)], the second band has Zak phase $\pi$ 
and the electrical field profiles at the $k=0$ and $k=\pi$ points have opposite parities. In contrast, for $d=0.75a$ [Fig.~1(d)], 
the second band has Zak phase 0, while the field profiles at the $k=0$ and $k=\pi$ points are
of the same parity. As envisioned by Zak~\cite{Zak1989}, the Zak phase is connected with the parity inversion according to
\beq
\frac{\theta}{\pi} = \frac{1}{2} [ \xi(k=0)-\xi(k=\pi) ] : {\rm mod} ~2 ,
\eeq
where $\xi(k=0)$ and $\xi(k=\pi)$ are the parity ($\pm 1$) at $k=0$ and $\pi$, respectively.
Our numerical calculation of the Zak phase through the Wilson-loop approach agrees with the parity inversion 
picture given by the above equation.

\subsection{\label{Chern calc}Chern number in 2D photonic crystals}

In Figs.~2(a) and 2(b) we illustrate how to slice the first BZ of 2D square (or rectangular) lattices and triangle (or 
honeycomb, kagome) lattices. For square or rectangular lattices, a closed loop in the BZ can be formed directly by running 
$k_x$ from $-\pi$ to $\pi$ for each fixed $k_y$. For triangle, honeycomb, and kagome lattices, we deform the first BZ from a 
hexagon to a rhomb. In this way, a closed loop is formed by running $k_2$ from $-\pi$ to $\pi$ for each given $k_1$. For both 
situations, we plot the Berry phase as a function of $k_x$ or $k_1$. The Chern number can be determined by examining the 
winding number of the Berry phase for $k_x\in [-\pi, \pi]$ or $k_1\in [-\pi, \pi]$. A black curve on the torus-shaped BZ is plotted in
Fig.~2(c) to represent the closed-loop with a fixed $k_x$ or $k_1$. Another red curve represents the  $k_x$ or $k_1$ 
direction along which we study the winding properties of the Berry phase to extract the topological numbers of the energy bands.

\begin{figure}[b]
	\includegraphics[width=3in]{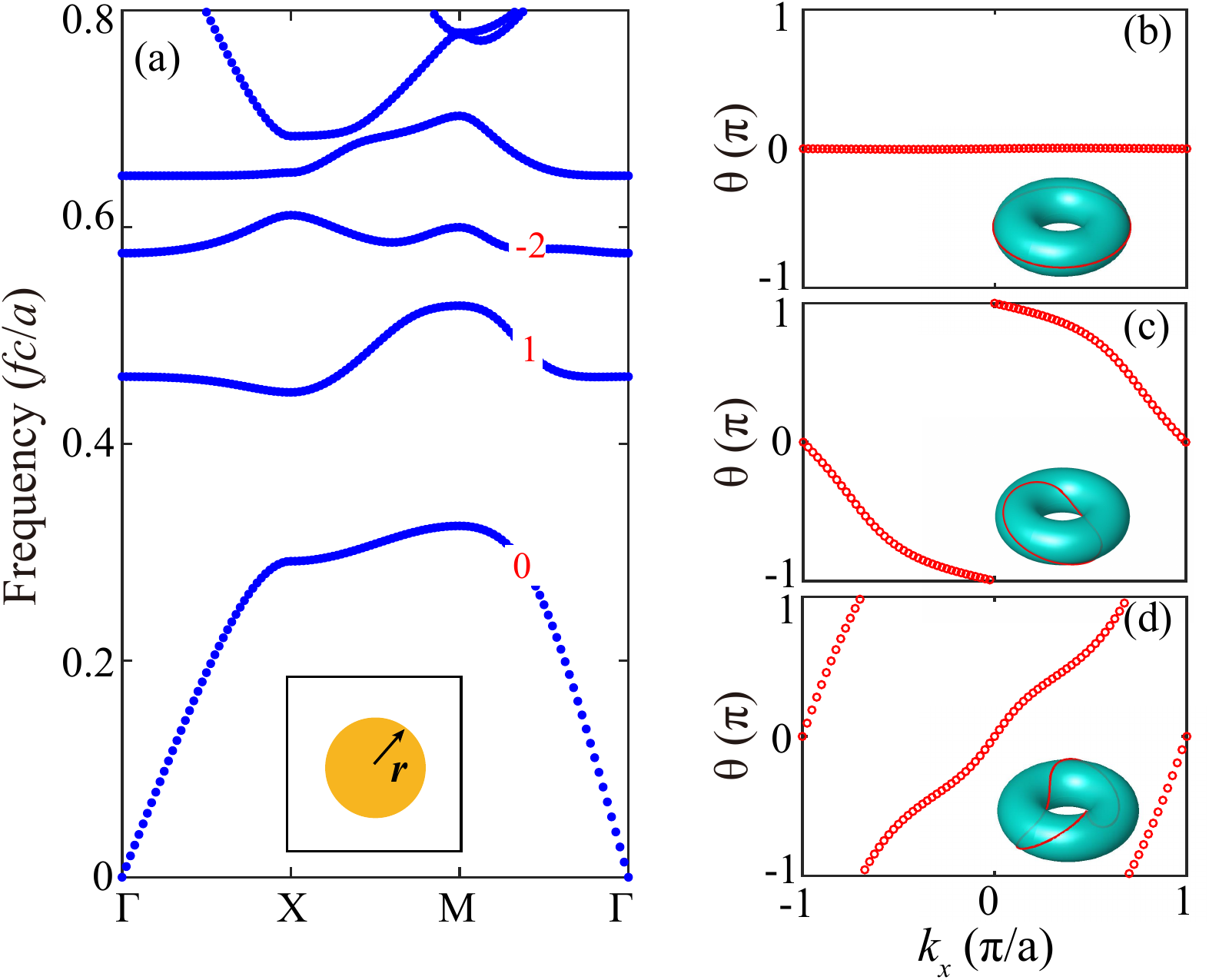}
	\caption{Numerical calculation of Chern number in 2D photonic crystal comprised of a square lattice of YIG rods with $\epsilon=15$ and $r=0.11a$. (a) Band structure of photonic crystal with applied magnetic field. The Chern number of each band is labeled in red. (b,c,d) The evolution of the Berry phase of (b) the first, (c) the second, (d) the third photonic band as the functions of $k_x$. The winding number of the Berry phase gives the Chern number.}
\end{figure}

\begin{figure}[b]
	\includegraphics[width=3in]{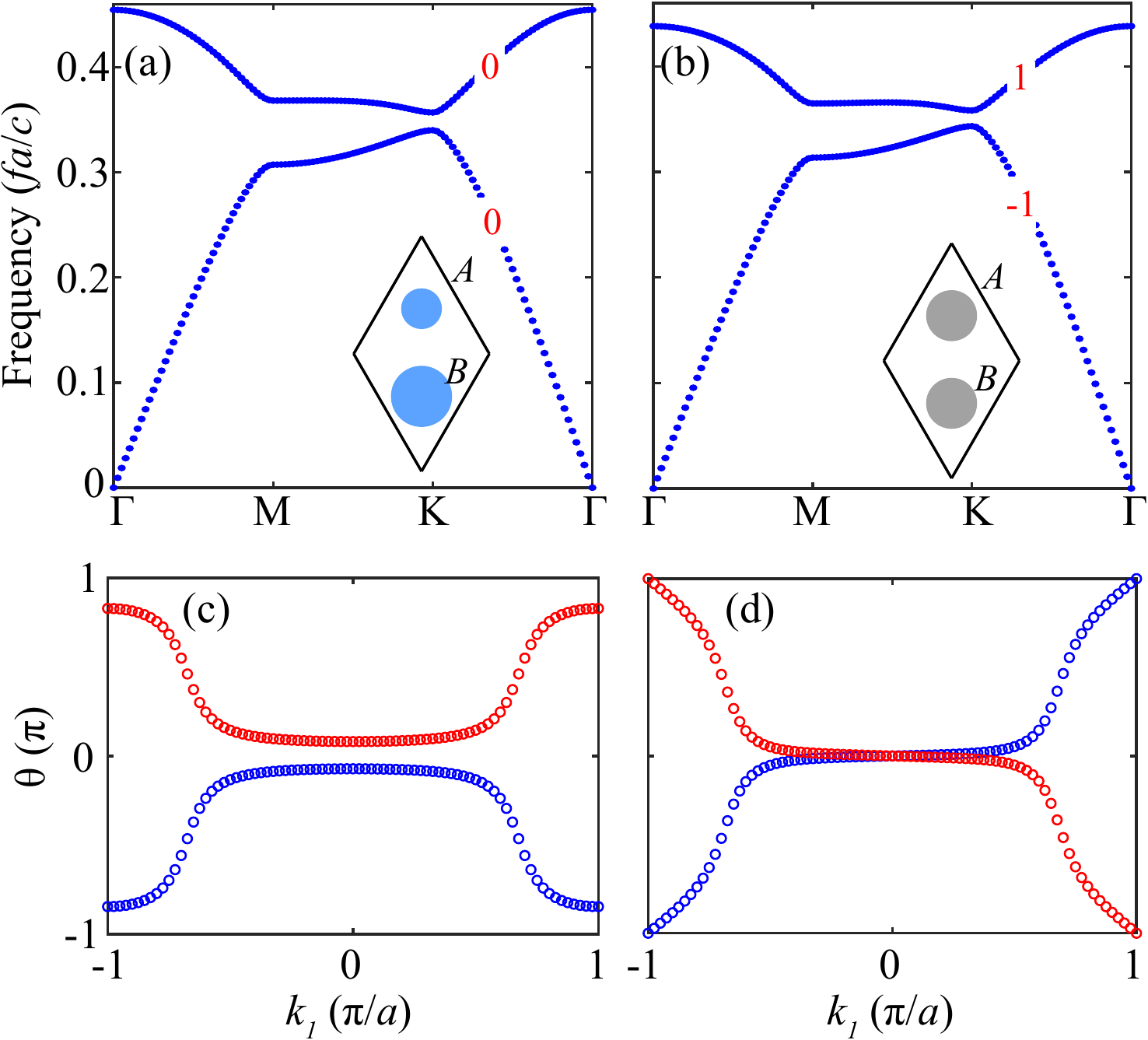}
	\caption{Numerical calculation of Chern number in honeycomb photonic crystal consist of YIG rods with $\epsilon=15$. (a,b) The band structure of honeycomb photonic crystal with (a) inversion symmetry breaking by making difference of two rods ($r_A=0.107a$, $r_B=0.115a$) and (b) time reversal symmetry breaking by applying external magnetic field ($r_A=r_B=0.125a$). (c,d) The evolution of the Berry phase for honeycomb photonic crystal with (c) inversion symmetry breaking and (d) time reversal symmetry breaking. In both (c) and (d), the blue and red points refer to the lower and upper bands, respectively.}
\end{figure}

To benchmark our calculation, we first consider a 2D photonic crystal comprised of a square lattice of yttrium-iron-garnet (YIG) 
rods for transverse-magnetic modes. This is the photonic crystal studied in Ref.~\onlinecite{c2}, which has photonic band gap 
with nontrivial Chern number. The material and geometry parameters are specified in the caption of Fig.~3. In order to attain 
nonzero Chern number, the time-reversal symmetry is broken by applying an external magnetic field along the $z$ direction. 
The magnetic field induces strong gyromagnetic anisotropy in YIG, leading to the following relative permeability\cite{c2},
\beq
\mu=\left[
\begin{matrix}
	15 & -12.4i & 0 \\
	12.4i & 15 & 0 \\
	0 & 0 & 1
\end{matrix}
\right].
\eeq
Essentially the permeability tensor is frequency-dependent in YIG materials. Nevertheless, it does not ruin the underlying physics, 
if the average permeability for the frequency-range of concern is adopted when the frequency-dependence is mild. The photonic 
band structures and Bloch wavefunctions are calculated using COMSOL Multiphysics, a commercial software for solving the Maxwell's
equations based on finite-element methods. The band structure is presented in Fig.~3(a). The Chern numbers of the first three bands are
determined by plotting the Berry phase $\theta_{n,k_x}$ $(n=1,2,3)$ as functions of $k_x$. Here the Berry phase is numerically
calculated from the electromagnetic field profiles by dividing the interval $k_y\in [-\pi, \pi]$ into a large 
number of small segments. For the first photonic band, we find that the Berry phase $\theta_{1,k_x}$ remains zero for all $k_x$, 
leading to zero Chern number [see Fig.~3(b)]. The Berry phase for the second photonic band $\theta_{2, k_x}$ winds from 0 to $2\pi$ 
when $k_x$ runs from $-\pi$ to $\pi$, revealing that the Chern number for the second band is ${\cal C}_2=1$ [Fig.~3(c)]. The Berry 
phase of the third band $\theta_{3,k_x}$ has a winding number -2, corresponding to a 
Chern number ${\cal C}_3=-2$, as shown in Fig.~3(d). The Chern numbers calculated here from the Wilson-loop approach agree
with the Chern numbers inferred from the chiral edge states in Ref.~\onlinecite{c2} using the bulk-edge correspondence~\cite{Hatsugai1993}.

We then apply the numerical Wilson-loop approach to honeycomb photonic crystals consisting of YIG rods. The honeycomb lattice has 
two rods in each unit cell. With identical rods and without breaking the time-reversal symmetry, there are Dirac points at the $K$ and 
$K^\prime$ points. Photonic band gap can be obtained by breaking either the inversion symmetry or the time-reversal symmetry 
[see Figs.~4(a) and 4(b)]. The former is realized by using two rods of different radii in each unit cell, while the latter is achieved by applying 
a magnetic field along the $z$ direction and break the time-reversal symmetry through magneto-optic effects. The results from the 
Wilson-loop approach are shown in Figs.~4(c) and 4(d). For the case with broken inversion symmetry, Fig.~4(c) indicates that the 
winding numbers of the Berry phase for the two photonic bands below and above the photonic band gap are zero, which indicates
vanishing Chern number. This situation is often denoted as the valley-Hall effect~\cite{va1,va2}, where the valley polarized Chern 
number for the $K$ and $K^\prime$ valleys are opposite and close to $\pm \frac{1}{2}$. Such valley Hall insulators can lead to
valley-polarized chiral edge states~\cite{va1,va2} [see also discussions in Sec.~IV of this paper]. For the case with broken 
time-reversal symmetry, Fig.~4(d) shows that the winding number of the Berry phase for the first photonic band 
gap is -1, thus the Chern number is ${\cal C}=-1$, agreeing with Haldane and Raghu's calculation in Ref.~\onlinecite{c1}.

\subsection{\label{spin-Chern calc} {Quantum spin Hall effect and fragile topology} in photonic and sonic crystals}
We now study the topological properties of quantum spin Hall effects in photonic and sonic crystals. Such photonic and sonic analogs of quantum spin Hall insulators have helical edge states on the boundary: a pair of edge states have opposite group velocities and angular momenta. 

In photonics, such phenomena can emerge in systems with ~\cite{Khanikaev2013,ma2015guiding,chen2014experimental,Wu2015,Xu2016} or without ~\cite{lu2016symmetry} time-reversal symmetry. In the literature, there are various proposals for photonic topological 
insulators~\cite{Khanikaev2013,hafezi2013imaging,ma2015guiding,chen2014experimental,Wu2015,Xu2016,Wang2016,he2016photonic,wang2017type}. 
For simplicity, we shall consider 2D all-dielectric photonic crystals for the transverse-magnetic modes, which can be realized by metallic plates cladding along the $z$ direction~\cite{ourexp}. Within this scheme, there are two ways to achieve photonic topological insulators. The first one is proposed in Ref.~\onlinecite{Wu2015} where the deformed honeycomb lattice photonic crystals are found to have helical edge states, resembling the quantum spin Hall effect of electrons. The second one is proposed in Ref.~\onlinecite{Xu2016} where topological phase transition is achieved by continuously tuning the inner and outer radii of core-shell triangle-lattice photonic crystals. 
The topological phase transition takes place at the accidental degeneracy between the $d$ and $p$ doublets which form a four-fold
degenerate Dirac point ~\cite{Xu2016}. Near the band degeneracy, an effective Hamiltonian for photons is constructed using the 
${\vec k}\cdot{\vec p}$ theory as derived from the Maxwell equations [see Refs.~\onlinecite{Xu2016,Wang2016,wang2017type}]. 
For both proposals, the pseudo-spins are emulated by the orbital angular momenta, which served as the photonic analogue of the 
electronic Kramers pairs, and the effective Hamiltonian for photons can be directly mapped to the Bernevig-Hughes-Zhang 
Hamiltonian~\cite{bhz} for the quantum spin Hall insulators of electrons. For acoustics, the sonic analog quantum spin Hall are 
realized using similar schemes, as proposed and realized in Refs.~\onlinecite{chen2014,He2016,acousticcoreshell}.

Below we present numerical calculation of the  {Berry phase} for acoustic topological insulators studied in
Refs.~\onlinecite{chen2014,He2016}. Specifically, the sonic crystal is made of a graphene-like lattice of steel rods. The acoustic 
band structures are presented in Figs.~5(a) and 5(b), representing a normal and topological acoustic band gaps, respectively. 
 {At the $\Gamma$ point}, the normal acoustic band gap has the $d$ bands above the $p$ bands, while the topological photonic band gap has  the $p$ bands above the $d$ bands. The transition between these two band gaps is realized by tuning the radii of the two steel rods (or the ``filling ratio"). With the increase of the radii of the two steel rods, the acoustic bands experience the processes of band gap closing and reopening, accompanying with the emergence of a four-fold degenerate Dirac point  {at the phase transition point}, as schematically depicted in the background of Figs.~5(a) and 5(b).

\begin{figure}
\includegraphics[width=3in]{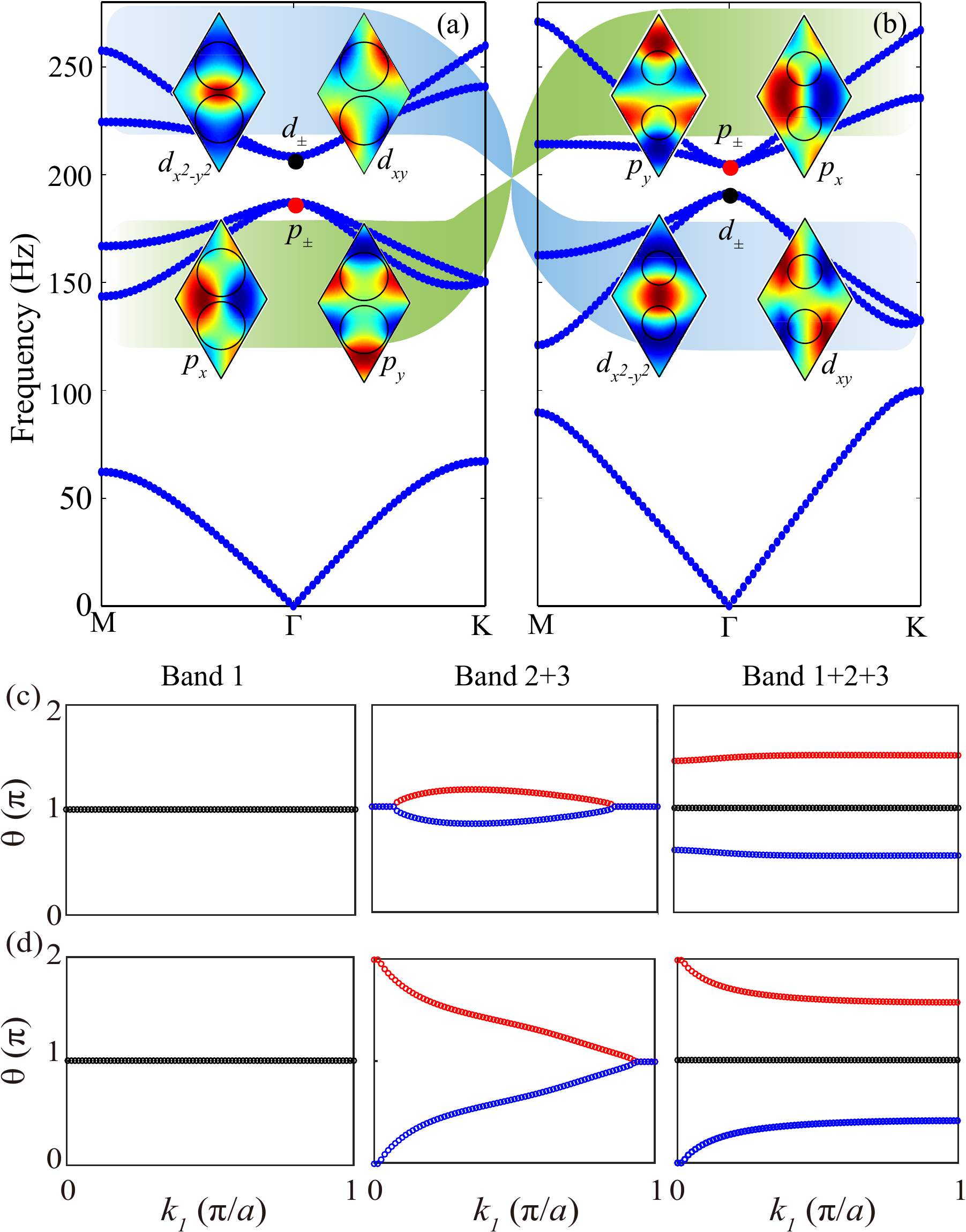}
\caption{Numerical calculation of  {Berry phase} in graphene-like sonic crystal. (a, b) Band structure of 
graphene-like sonic crystal with (a)  {trivial insulator} case, $r=0.3a$, (b)  {quantum spin Hall effect} case, $r=0.45a$. The insets show the Bloch states at the BZ center corresponding to different parities, which indicate parity inversion. (c, d) The evolution of Berry phase with different sets of bands for (c) the normal insulating phase, ${\cal C}_s=0$ and (d) the quantum spin Hall insulator phase, ${\cal C}_s=1$. Left panel: first band. Middle panel: first and second bands. Right panel: lowest three bands. The material parameters used for calculation are the density and longitudinal sound speed (stainless steel: 7880~kg/m$^3$ and 6010~m/s , air: 1.25~kg/m$^3$ and 343~m/s).}
\end{figure}

Using the Bloch wavefunctions from first-principle calculations and adopting the Wilson-loop approach, we calculate the Berry phases for each given $k_1$, and examine the winding properties of the Berry phases. There are three acoustic bands below the second band gap of which the topological property is studied here. Note that the first acoustic band is isolated from the second and third acoustic bands.  {We can calculate the Wannier band for the first acoustic band [see the left panels of 
Figs.~5(c) and 5(d)]. The calculation results show that the first acoustic band has Berry phase $\theta=\pi$ for all $k_1$ for both
the normal and topological band gaps. This Berry phase has no connection with the quantum spin Hall effect. The Berry phase
characteristics of the acoustic quantum spin Hall effect can be revealed by the Wilson-loop calculation for the second and third
bands. The Wannier bands for this set of bands is shown in the middle panels of Figs.~5(c) and 5(d). It is seen that the 
Wannier bands in the region $k_1\in [0,\pi]$ are gapless for the quantum spin Hall insulator, whereas the Wannier bands
are gapped for the normal band gap. Here, by the term ``gapless" it means that the Wannier bands spread over the whole
$[0, 2\pi]$ region, as defined in Ref.~\cite{YuRui}. The Berry phase characterization is thus consistent with the discovery in Refs.~\onlinecite{chen2014,He2016}.}

However, since the pseudo-spin degeneracy in acoustic systems is not guaranteed by time-reversal symmetry 
(because phonons are bosons) but by crystalline symmetries, the acoustic quantum spin Hall insulator is not as 
robust as the electronic quantum spin Hall insulator where the band topology is protected by the global time-reversal 
symmetry.  {Moreover, in contrast to the electronic quantum spin Hall systems, the
photonic/acoustic quantum spin Hall systems feature fragile topology. In the literature, bands with fragile topology
have gapless Wannier bands by themselves, but the Wannier bands can become gapped when a set of bands
with gapped Wannier bands are added to the Wilson-loop calculation. Indeed, when the first band is added to the
Wilson-loop calculation (i.e., the calculation include the first three bands), the Wannier bands become gapped,
as shown in the right panels of Figs.~5(c) and 5(d). These results confirm that the acoustic system studied here
has fragile topology. In fact, classical wave systems (e.g., sonic crystals and photonic crystals~\cite{}) are ideal
platform for the study of fragile topology. These topological systems have gapped edge states, unless certain edge 
symmetry protect them~\cite{}.  }

\section{Bulk-edge correspondence}
\label{sec:level4}
\subsection{Chern number and chiral edge state}
The bulk-edge correspondence is of fundamental importance in topological physics, which gives the relationship between 
the total number of edge states in a specific band gap and the topological properties of all the bulk bands below the band gap. 
In this section, we check the Wilson-loop calculation of the topological number by studying the edge states according to 
the bulk-edge correspondence in photonic/acoustic topological systems.
	
In many situations, the edge states in the nontrivial topological systems can be understood through the seminal 
Jackiw-Rebbi theory~\cite{Jackiw} which studies a continuum model of a 1D massive Dirac equation with a domain 
wall where the Dirac mass switches sign. The solution of such a position-dependent massive Dirac equation is a 
topological mid-gap mode which is localized at the domain wall and decays exponentially away from the domain wall
[see Fig.~6(a)]. The 2D extension of the Jackiw-Rebbi theory gives chiral edge states for the Haldane model
with Chern number ${\cal C}=\pm 1$~\cite{Hasan2010}, since the topological phase transition in the Haldane model
between the valley Hall insulator with ${\cal C}=0$ to the Chern insulator with ${\cal C}=\pm 1$ can be understood 
as the sign change of the Dirac mass at one of the valley ($K$ or $K^\prime$)~\cite{Haldane1988}.
	
We first give an example of the Jackiw-Rebbi bound state in a 1D photonic crystal. The topological transition of 1D binary 
photonic crystal is featured with band gap closing (with a Dirac point) and reopening. Such a topological transition can also be 
understood as the sign change of the Dirac mass in the massive Dirac equation. When two photonic crystals with
opposite Dirac masses are placed together, a bound state emerge at the interface between them.
To construct such a domain wall, we place two 1D binary photonic crystals with $d=0.1a$ and $d=0.75a$ [see Fig.~1] together.
They have a common frequency gap from $0.79\tfrac{2\pi c}{a}$ to $0.91\tfrac{2\pi c}{a}$. The total Zak phases of the
photonic bands below the common frequency gap for the two photonic crystals are $0$ and $\pi$, respectively. The 
photonic spectrum of the combined photonic crystals is displayed in Fig.~6(b). There are two degenerate mid-gap bound 
states (red dots), of which the electromagnetic fields are localized at the interface [see the inset].
	
To study the chiral edge states in 2D systems, we use the honeycomb-lattice photonic crystals that have been studied in 
Sec.~\ref{Chern calc} which break the inversion symmetry or time-reversal symmetry, respectively. We then calculate the edge
and bulk photonic spectrum using a supercell with a zigzag boundary along the $x$-direction. The zigzag boundary is formed
by the photonic crystal and a perfect-electric-conducting boundary. The detailed parameters of the
photonic crystals are the same as for Fig.~4. The calculated photonic spectrum of the supercells are shown in Figs.~6(c) and 6(d). 
For the photonic crystal with Chern number ${\cal C}=0$, there are no  {gapless} edge states in the photonic band gap. For the photonic
crystal with Chern number ${\cal C}=1$, there are chiral edge states (red color) in the potonic band gap. These results 
are consistent with the calculated Chern number and the bulk-edge correspondence principle.

\begin{figure}
\includegraphics[width=3in]{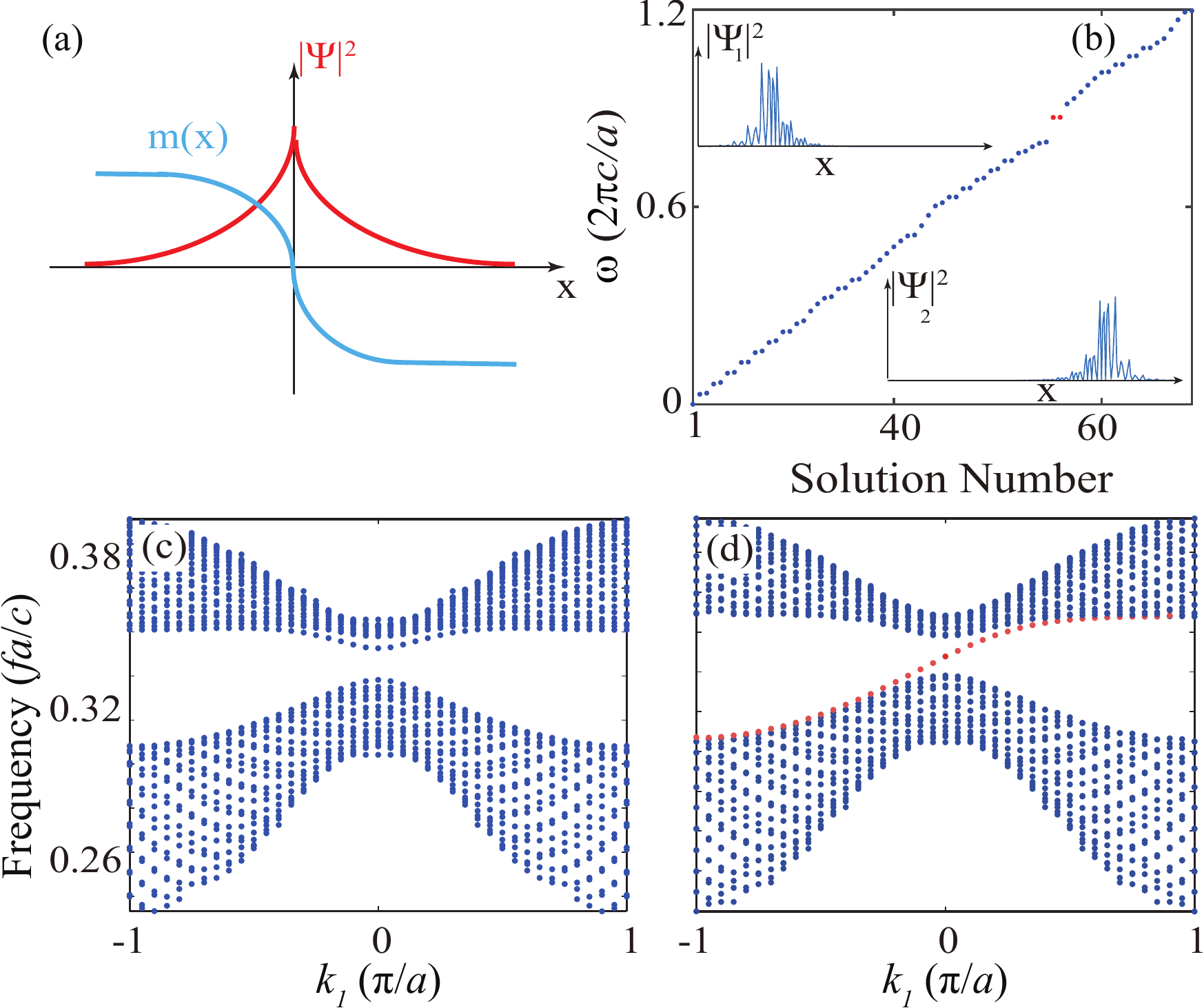}
\caption{Topological mid-gap states and chiral edge states. (a) Schematic of the topological mid-gap mode. The distribution of Dirac mass and topological mid-gap mode are represented by blue and red lines, respectively. (b) Eigen-frequency solution of the combined photonic crystal composed of 1D photonic crystal studied in Fig. 1. Inset: topological mid-gap mode. (c, d) The projected band structure mentioned in Fig. 2 with (c) inversion symmetry breaking and (d) time-reversal symmetry breaking. Both cases are calculated with zigzag boundary.}
\end{figure}

\subsection{ {Quantum spin Hall effect} and helical edge state}
Helical edge states, the hallmark of quantum spin Hall effects, have been widely studied in photonic/acoustic 
system~\cite{Khanikaev2013,ma2015guiding,chen2014experimental,Wu2015,Xu2016,He2016}. Here we perform 
calculations for the edge states using the sonic crystals studied in Sec.~\ref{spin-Chern calc}. As shown in Fig.~7(a), 
there are a pair of edge states in the acoustic band gap. We further plot the acoustic pressure profiles (both the 
amplitude and the phase) for the two edge states at $k_1=0.1\frac{\pi}{a}$ (labeled as A and B) in Fig.~7(b). The 
acoustic pressure phase profiles of the edge states A and B indicate clockwise and anti-clockwise winding phase vortices, 
as indicated by the green arrows in the left panels of Fig.~7(b). These phase vortices indicate finite and opposite orbital 
angular momenta, which mimic the opposite spin angular momentum in the helical edge states in the electronic quantum 
spin Hall insulator.
	
One of the difference between the electronic quantum spin Hall insulators and the bosonic or classical analogs of 
quantum spin Hall insulators is that the former is a topological order protected by the global time-reversal symmetry, whereas
the latter is a topological crystalline order protected by crystalline symmetries. For electronic and other fermionic systems,
time-reversal symmetry protects the Kramers degeneracy and the crossing of the helical edge states at time-reversal invariant
wavevectors. However, for bosonic and classical systems, the pseudo-Kramers degeneracy is induced by the pseudo-time-reversal 
symmetry induced by the crystalline symmetry. For the sonic crystals studied here the $C_6$ rotation symmetry inducing
the pseudo-Kramers degeneracy is broken on the edge boundary. Therefore, there is a tiny band gap in the dispersion of the 
helical edge states [almost unnoticeable in Fig.~7(a)]. In general the crossing between the two branches of the helical
edge states are not protected when the boundary does not have the crystalline symmetry that protects the pseudo-Kramers
degeneracy,  {which is often true for crystals with symmorphic symmetry~\cite{Wu2015,Xu2016,deformedPhC,chen2014,He2016}.
However, in nonsymmorphic symmetry crystals the gapless edge states can be protected by glide symmetries
which are preserved on the edges, as shown in Refs.~\cite{lu2016symmetry,Jiang2018}. The possibility that the edge states
may open up a gap even when the bulk topological properties are well-defined is a key feature of fragile topology. This feature
may reduce the topological robustness of the edge states, making them less efficient for one-way transport.}

\begin{figure}
\includegraphics[width=3.5in]{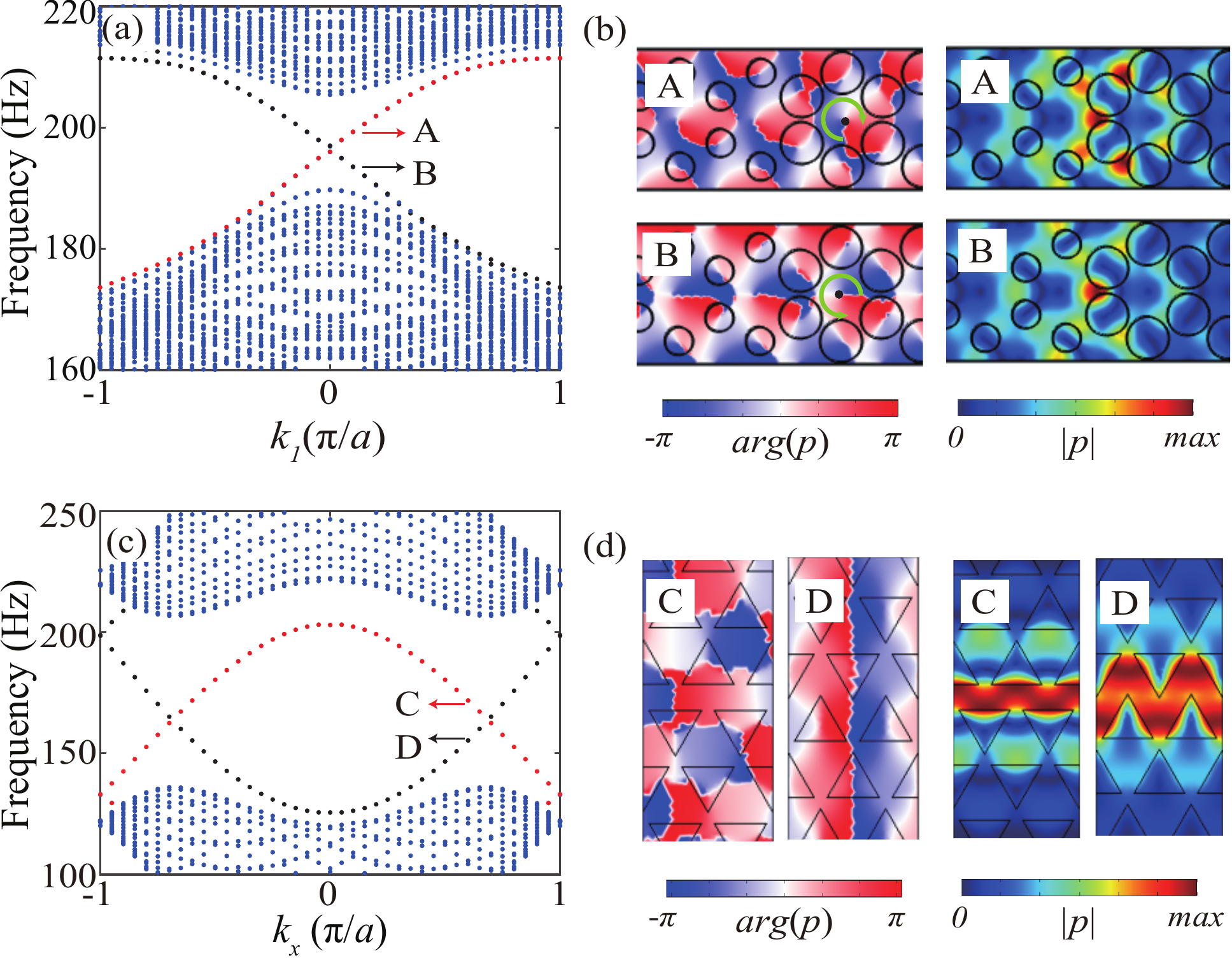}
\caption{Edge states in acoustic topological insulator and sonic crystal with valley Hall insulting phases. (a) The projected bulk band (blue points) and helical edge states (black and red points) in the acoustic topological insulator. (b)The phase distribution (right panel) and acoustic pressure profile $|p|$ (left panel) of states A and B labeled in (a). (c) The projected bulk band (blue points) and valley-chiral edge states (black and red points) in the sonic crystal with valley Hall insulating phases.(d) The phase distribution (left panel) and acoustic pressure profile $|p|$ (right panel) of states C and D labeled in (c). The side length of triangular-shaped rod is $0.8a$. Other material parameters used for calculation are the same as those in Sec.~\ref{spin-Chern calc}.}
\end{figure}
	
We further elaborate the bulk-edge correspondence by pointing out a difference between the edge states of the quantum 
spin Hall insulator and those of the valley Hall insulator (both of them for acoustic systems). The acoustic valley Hall insulators
studied here are adopted from Refs.~\onlinecite{va1,va2} which are realized by sonic crystals made by hexagonal array of 
triangular rod-like scatters. It has been shown that the Dirac points at the $K$ and $K^\prime$ points can be gapped by 
rotating the triangular scatters, accompanying with the symmetry reduction from the $C_{3v}$ point group to the $C_{3}$ 
point group. The resultant valley Hall insulators have opposite valley Chern numbers at the $K$ and $K^\prime$ points. 
Both the $K$ and $K^\prime$ valleys can be described by the massive Dirac equations in two dimensions with Dirac masses. 
When two acoustic valley Hall insulators with a common band gap but opposite Dirac masses at the $K$ point are placed 
together to form a Dirac mass domain wall, there are valley-polarized chiral edge states~\cite{va1,va2,Mavalley}. The edge 
states at the $K^\prime$ valley is related to those at the $K$ valley as dictated by the time-reversal symmetry [see Fig.~7(c)]. 
These edge states with opposite group velocities seem to be phenomenologically like the helical edge states for the 
quantum spin Hall insulators, particularly when armchair boundaries are considered~\cite{armchair}. 

However, there is a difference between the edge states for the acoustic quantum spin Hall insulators and those for the acoustic
valley Hall insulators: the former have phase vortices (indicating finite orbital angular momenta) while the latter do not 
have phase vortex (or have multiple phase vortices with opposite winding numbers, resulting vanishing orbital angular 
momentum) as indicated in Fig.~7(d). Such a difference indicates that the edge states for the acoustic valley Hall 
insulators cannot be regarded as pseudo-spin polarized helical edge states, reflecting that the bulk bands for those 
two types of insulators have distinct topological properties. Indeed, as shown in Fig.~4, Wannier bands are gapped
for the quantum valley Hall case, whereas the Wannier bands are gapless for the quantum spin Hall case.

\section{Conclusion}
\label{sec:level5}
In conclusion, we presented the Wilson-loop approach toward topological numbers for classical (i.e., photonic and acoustic) 
energy bands. Specifically, based on the Wilson-loop approach, we obtained the  {Zak phase} and Chern number for 1D and 2D photonic bands from first-principle calculations, starting from the definition of Berry phases in those classical systems. Such first-principle calculations rigorously verify the topological numbers in several topological classical systems widely studied in the literature. We further elaborate on the subtle features of the edge states in those topological classical-wave systems---in particular, the phase vortices in the edge states of acoustic quantum spin Hall insulators and the absence of such vortices for acoustic quantum valley Hall insulators. Moreover, using the Wilson-loop approach we are able to reveal the nature of fragile topology of those topological classical-wave systems, namely that
the nontrivial winding of the Berry phases can be destroyed by adding trivial bulk bands. Those aspects, as complimentary
to the existing studies, enriches the understanding of the underlying physics of topological classical-wave systems.\\

\begin{acknowledgments}
J.-H.J and H.-X.W acknowledge supports from the Jiangsu distinguished professor funding and the National Science Foundation of China (Grant No.11675116) and the Soochow University. G.-Y. G acknowledge supports from the Ministry of Science and Technology, the National Center for Theoretical Sciences and the Thematic Research Program (AS-TP-106-M07) of Academia Sinica in Taiwan. J.-H.J and H.-X.W thanks Hongming Weng, Chunyin Qiu and Zhengyou Liu for helpful and illuminating discussions. 
\end{acknowledgments}


\begin{thebibliography}{10}

\bibitem{Thouless1982}
D. J. Thouless, M. Kohmoto, M. P. Nightingale, and	M. den Nijs, 
\newblock {\it Phys. Rev. Lett.} {\bf 49}, 405(1982).

\bibitem{KOHMOTO1985343}
M. Kohmoto, 
\newblock {\it Ann. Phys. } {\bf 160}, 343 (1985).

\bibitem{Haldane1988}
F.~D.~M. Haldane, 
\newblock {\it Phys. Rev. Lett.} {\bf 61}, 2015 (1988).

\bibitem{Hasan2010}
M.~Z. Hasan and C.~L. Kane, 
\newblock {\it Rev. Mod. Phys.} {\bf 82}, 304 (2010).

\bibitem{Qi2011}
X.-L. Qi and S.-C. Zhang, 
\newblock {\it Rev. Mod. Phys.} {\bf 83}, 1057 (2011).

\bibitem{tsm1}
O. Vafek and A. Vishwanath, 
\newblock {\it Ann. Rev. Cond. Matt. Phys.} {\bf 5}, 83 (2014).

\bibitem{kanemele}
C.~L. Kane and E.~J. Mele, 
\newblock {\it Phys. Rev. Lett.} {\bf 95}, 146802 (2005).

\bibitem{bhz}
B. A. Bernevig, T. L. Hughes, and S.-C. Zhang, 
\newblock {\it Science} {\bf 314}, 1757 (2006).

\bibitem{konig}
M. K{\"o}nig, S. Wiedmann, C. Br{\"u}ne, A. Roth, H.
Buhmann, L.~W. Molenkamp, X.-L. Qi and S.-C. Zhang, 
\newblock {\it Science} {\bf 318}, 766 (2007).

\bibitem{fukane1}
L. Fu, C.~L. Kane, and E.~J. Mele, 
\newblock {\it Phys. Rev. Lett.} {\bf 98}, 106803 (2007).

\bibitem{fukane2}
L. Fu and C.~L. Kane, 
\newblock {\it Phys. Rev. B} {\bf 76}, 045302 (2007).

\bibitem{fukane3}
L. Fu and C.~L. Kane, 
\newblock {\it Phys. Rev. Lett.} {\bf 100}, 096407 (2008).

\bibitem{ckxu}
C. Xu and J.~E. Moore, 
\newblock {\it Phys. Rev. B} {\bf 73}, 045322 (2006).

\bibitem{jiang2}
J.-H. Jiang and S.~Wu, 
\newblock {\it Phys. Rev. B} {\bf 83}, 205124 (2011).

\bibitem{murakami}
S. Murakami, 
\newblock {\it New J. Phys.} {\bf 9}(9), 356 (2007).

\bibitem{burkov2011weyl}
A.~A. Burkov and L. Balents, 
\newblock {\it Phys. Rev. Lett.} {\bf 107}, 127205 (2011).

\bibitem{jiang1}
J.-H. Jiang, 
\newblock {\it Phys. Rev. A} {\bf 85}, 033640 (2012).

\bibitem{luling1}
L. Lu, Z. Wang, D. Ye, L. Ran, L. Fu, J. D. Joannopoulos, and
M. Solja{\v{c}}i{\'c}, 
\newblock {\it Science} {\bf 349}, 622 (2015).

\bibitem{c1}
F.~D.~M. Haldane and S.~Raghu, 
\newblock {\it Phys. Rev. Lett.} {\bf 100}, 013904 (2008).

\bibitem{c2}
Z. Wang, Y. D. Chong, J. D. Joannopoulos, and M. Solja{\v{c}}i{\'c}, 
\newblock {\it Phys. Rev. Lett.} {\bf 100}, 013905 (2008).

\bibitem{c3}
Z. Wang, Y. D. Chong, J. D. Joannopoulos, and M. Solja{\v{c}}i{\'c}, 
\newblock {\it Nature} {\bf 461}(7265), 772 (2009).

\bibitem{c4}
X. Ao, Z. Lin, and C. T. Chan, 
\newblock {\it Phys. Rev. B} {\bf 80}, 033105 (2009).

\bibitem{poo2011}
Y. Poo, R.-X. Wu, Z. Lin, Y. Yang, and C. T. Chan, 
\newblock {\it Phys. Rev. Lett.} {\bf 106}, 093903 (2011).

\bibitem{hafezi2013imaging}
M. Hafezi, S. Mittal, J. Fan, A. Migdall, and J. Taylor, 
\newblock {\it Nat. Photon.} {\bf 7}, 1001 (2013).

\bibitem{Mittal2016}
S. Mittal, S. Ganeshan, J. Fan, A. Vaezi, and M. Hafezi, 
\newblock {\it Nat. Photon.} {\bf 10}, 180 (2016).

\bibitem{rechtsman2013photonic}
M. C. Rechtsman, J. M. Zeuner, Y. Plotnik, Y. Lumer,
D. Podolsky, F. Dreisow, S. Nolte, M. Segev, and A. Szameit, 
\newblock {\it Nature} {\bf 496}, 196 (2013).

\bibitem{Khanikaev2013}
A. B. Khanikaev, S. Hossein Mousavi, W.-K. Tse, M. Kargarian,
A. H. MacDonald, and G. Shvets, 
\newblock {\it Nat.Mater.} {\bf 12}, 233 (2013).

\bibitem{ma2015guiding}
T. Ma, A. B. Khanikaev, S. H. Mousavi, and G. Shvets, 
\newblock {\it Phys. Rev. Lett.} {\bf 114}, 127401 (2015).

\bibitem{chen2014experimental}
W.-J. Chen, S.-J. Jiang, X.-D. Chen, B. Zhu, L. Zhou, J.-
W. Dong, and C. T. Chan, 
\newblock {\it Nat. Commun.} {\bf 5}, 5782 (2014).

\bibitem{Wu2015}
L.-H. Wu and X. Hu, 
\newblock {\it Phys. Rev. Lett.} {\bf 114}, 223901 (2015).

\bibitem{Xu2016}
L. Xu, H.-X. Wang, Y.-D. Xu, H.-Y. Chen, and J.-H.
Jiang, 
\newblock {\it Opt. Express} {\bf 24}, 18059 (2016).

\bibitem{deformedPhC}
X. Zhu, H.-X. Wang, C. Xu, Y. Lai, J.-H. Jiang, and
S. John, 
\newblock {\it Phys. Rev. B} {\bf 97}, 085148 (2018).

\bibitem{Wang2016}
H.-X. Wang, L. Xu, H. Chen, and J.-H. Jiang,
\newblock {\it Phys. Rev. B} {\bf 93}, 235155 (2016).

\bibitem{c10}
S. A. Skirlo, L. Lu, and M. Solja{\v{c}}i{\'c}, 
\newblock {\it Phys. Rev. Lett.} {\bf 113}, 113904 (2014).

\bibitem{c11}
D. Leykam, M. C. Rechtsman, and Y. D. Chong,
\newblock {\it Phys. Rev. Lett.} {\bf 117}, 013902 (2016).

\bibitem{c12}
F. Gao, Z. Gao, X. Shi, Z. Yang, X. Lin, H. Xu, J.~D.
Joannopoulos, M. Solja{\v{c}}i{\'c}, H. Chen, L. Lu, Y. Chong, and
B. Zhang, 
\newblock {\it Nat. Commun.} {\bf 7}, 11619 (2016).

\bibitem{z2new}
X. Cheng, C. Jouvaud, X. Ni, S. H. Mousavi, A. Z. Genack,
and A. B. Khanikaev, 
\newblock {\it Nat. Mater.} {\bf 15}, 542 (2016).

\bibitem{wang2017type}
H.-X. Wang, Y. Chen, Z. H. Hang, H.-Y. Kee, and J.-H.
Jiang, 
\newblock {\it npj Quantum Materials} {\bf 2}, 54 (2017).

\bibitem{Lu2014}
L. Lu, J.~D. Joannopoulos, and M. Solja{\v{c}}i{\'c}, 
\newblock {\it Nat. Photon.} {\bf 8}, 821 (2014).

\bibitem{Lu2016}
L. Lu, J.~D. Joannopoulos, and M. Solja{\v{c}}i{\'c}, 
\newblock {\it Nat. Phys.} {\bf 12}, 626 (2016).

\bibitem{RMPtopophoton}
T. Ozawa, H. M. Price, A. Amo, N. Goldman, M. Hafezi, L. Lu, M. C. Rechtsman, D. Schuster, J. Simon, O. Zilberberg, and I. Carusotto, 
\newblock {\it Rev. Mod. Phys.} {\bf 1}, 015006 (2019).

\bibitem{Yang2015topoacousitc}
Z. Yang, F. Gao, X. Shi, X. Lin, Z. Gao, Y. Chong, and
B. Zhang, 
\newblock {\it Phys. Rev. Lett.} {\bf 114}, 114301 (2015).

\bibitem{Xiao2015}
M. Xiao, W.-J. Chen, W.-Y. He, and C. T. Chan, 
\newblock {\it Nat.Phys.} {\bf 11}, 920 (2015).

\bibitem{He2016}
C. He, X. Ni, H. Ge, X.-C. Sun, Y.-B. Chen, M.-H. Lu,
X.-P. Liu, and Y.-F. Chen, 
\newblock {\it Nat.Phys.} {\bf 12}, 1124 (2016).

\bibitem{chen2014}
Z.-G. Chen, X. Ni, Y. Wu, C. He, X.-C. Sun, L.-Y. Zheng,
M.-H. Lu, and Y.-F. Chen, 
\newblock {\it Sci. Rep.} {\bf 4}, 4613 (2014).

\bibitem{wuying}
Z.-G. Chen and Y. Wu, 
\newblock {\it Phys. Rev. Applied} {\bf 5}, 054021 (2016).

\bibitem{va1}
J. Lu, C. Qiu, M. Ke, and Z. Liu, 
\newblock {\it Phys. Rev. Lett.} {\bf 116}, 093901 (2016).

\bibitem{va2}
J. Lu, C. Qiu, L. Ye, X. Fan, M. Ke, F. Zhang, and Z. Liu, 
\newblock {\it Nat. Phys.} {\bf 13}, 369 (2016).

\bibitem{liu3}
J. Lu, C. Qiu, S. Xu, Y. Ye, M. Ke, and Z. Liu, 
\newblock {\it Phys. Rev. B} {\bf 89}, 134302 (2014).

\bibitem{zhu1}
Y. Ding, Y. Peng, Y. Zhu, X. Fan, J. Yang, B. Liang, X. Zhu, X. Wan, and J. Cheng,
\newblock {\it Phys. Rev. Lett.} {\bf 122}, 014302 (2019).

\bibitem{zhu2}
Y.-G. Peng, C.-Z Qin, D.-G. Zhao, Y.-X. Shen, X.-Y. Xu, M. Bao, H. Jia, and X.-F. Zhu
\newblock {\it Nat. Commun.} {\bf 7}, 13368 (2016).

\bibitem{Zhang2017}
Z. Zhang, Q.Wei, Y. Cheng, T. Zhang, D.Wu, and X. Liu, 
\newblock {\it Phys. Rev. Lett.} {\bf 118}, 084303 (2017).

\bibitem{Jiang2018}
X. Zhang, H.-X. Wang, Z.-K. Lin, Y. Tian, B. Xie, M.-H. Lu, 
Y.-F. Chen, and J.-H. Jiang, 
\newblock {\it Nat. Phys.} {\bf 15}, 582 (2019).

\bibitem{mechanical1}
R. Süsstrunk and S. D. Huber,
\newblock {\it Science} {\bf 349}, 47 (2015).

\bibitem{mechanical2}
J. Vila, R. K. Pal, and M. Ruzzene, 
\newblock {\it Phys. Rev. B} {\bf 96}, 134307 (2017).

\bibitem{mechanical3}
D. Z. Rocklin, B. G. Chen, M. Falk, V. Vitelli, and T. C.
Lubensky,
\newblock {\it Phys. Rev. Lett.} {\bf 116}, 135503 (2016).

\bibitem{z2linenode}
Z. Xiong, H.-X. Wang, H. Ge, J. Shi, J. Luo, Y. Lai, M.-H.
Lu, and J.-H. Jiang
\newblock {\it Phys. Rev. B} {\bf 97}, 180101 (2018).

\bibitem{Onoda2006}
M. Onoda, S. Murakami, and N. Nagaosa, 
\newblock {\it Phys. Rev. E} {\bf 74}, 066610 (2006).

\bibitem{z21}
M.~G. Silveirinha, 
\newblock {\it Phys. Rev. B} {\bf 93}, 075110 (2016).

\bibitem{Zak1989}
J.~Zak, 
\newblock {\it Phys. Rev. Lett.} {\bf 62}, 2747 (1989).

\bibitem{hmwengreview}
H. Weng, R. Yu, X. Hu, X. Dai, and Z. Fang, 
\newblock {\it Adv. Phys.} {\bf 64}, 227 (2015).

\bibitem{lu2016symmetry}
L. Lu, C. Fang, L. Fu, S. G. Johnson, J. D. Joannopoulos, 
and M. Solja{\v{c}}i{\'c}, 
\newblock {\it Nat. Phys.} {\bf 12}, 337 (2016).

\bibitem{Hatsugai1993}
Y. Hatsugai, 
\newblock {\it Phys. Rev. Lett.} {\bf 71}, 3697 (1993).

\bibitem{he2016photonic}
C. He, X.-C. Sun, X.-P. Liu, M.-H. Lu, Y. Chen, L. Feng,
and Y.-F. Chen, 
\newblock {\it Proc. Natl. Acad. Sci. USA} {\bf 113}, (2016).

\bibitem{ourexp}
Y. Yang, Y. F. Xu, T. Xu, H.-X.Wang, J.-H. Jiang, X. Hu,
and Z. H. Hang, 
\newblock {\it Phys. Rev. Lett.} {\bf 120}, 217401 (2018).

\bibitem{acousticcoreshell}
J. Mei, Z. Chen, and Y. Wu, 
\newblock {\it Sci. Rep.} {\bf 6}, 32752 (2016).

\bibitem{Fragile}
H. C. Po, H. Watanabe, and A. Vishwanath, 
\newblock {\it Phys. Rev. Lett.} {\bf 121}, 126402 (2018).

\bibitem{Jackiw}
R.~Jackiw and C.~Rebbi, 
\newblock {\it Phys. Rev. D} {\bf 13}, 3398 (1976).

\bibitem{Mavalley}
T. Ma and G. Shvets, 
\newblock {\it Phys. Rev. B} {\bf 95}, 165102 (2017).

\bibitem{armchair}
L. Zhang, Y. Yang, M. He, H.-X. Wang, Z. Yang, E. Li,
F. Gao, B. Zhang, R. Singh, J.-H. Jiang, and H. Chen, 
\newblock {arXiv preprint} arXiv:1805.03954 (2018).

\bibitem{YuRui}
R. Yu, X. Qi, A. Bernevig, Z. Fang, and X. Dai,
\newblock {\it Phys. Rev. B} {\bf 84}, 075119 (2011).

\bibitem{dipolemoments}
W. Benalcazar, A. Bernevig, and T. Hughes,
\newblock {\it Phys. Rev. B} {\bf 96}, 245115 (2017).

\bibitem{dipolemoments}
M. Paz, M. Vergniory, D. Bercioux, A. -Etxarri, and B. Bradlyn,
\newblock {arXiv preprint} arXiv:1903.02562 (2019).

\end{thebibliography}


\bibliographystyle{unsrt}


\end{document}